\documentclass[10pt,twocolumn,letterpaper]{article}

\usepackage{iccv}
\usepackage{times}
\usepackage{epsfig}
\usepackage{graphicx}
\usepackage{amsmath}
\usepackage{amssymb}
\usepackage{euscript}

\newcommand\blfootnote[1]{%
  \begingroup
  \renewcommand\thefootnote{}\footnote{#1}%
  \addtocounter{footnote}{-1}%
  \endgroup
}

\usepackage[pagebackref=true,breaklinks=true,letterpaper=true,colorlinks,bookmarks=false]{hyperref}

\iccvfinalcopy % *** Uncomment this line for the final submission

\def\iccvPaperID{10580} % *** Enter the ICCV Paper ID here

\ificcvfinal\fi

\begin{document}

%%%%%%%%% TITLE
\title{Attention-based Stylisation for Exemplar Image Colourisation}

\author{Marc Gorriz Blanch$^{\star}$$^{\dagger}$, Issa Khalifeh$^{\star}{\ddagger}$, Alan Smeaton$^{\dagger}$, Noel E. O’Connor$^{\dagger}$, Marta Mrak$^{\star}$ \\\\
$^{\star}$British Broadcasting Corporation\\
$^{\dagger}$Insight Centre for Data Analytics \& Dublin City University\\
$^{\ddagger}$Queen Mary University of London

}

\maketitle
\ificcvfinal\thispagestyle{empty}\fi

%%%%%%%%% ABSTRACT
\begin{abstract}
Exemplar-based colourisation aims to add plausible colours to a grayscale image using the guidance of a colour reference image. Most existing methods tackle the task as a style transfer problem, using a convolutional neural network (CNN) to obtain deep representations of the content of both inputs. Stylised outputs are then obtained by computing similarities between both feature representations in order to transfer the style of the reference to the content of the target input. However, in order to gain robustness towards dissimilar references, the stylised outputs need to be refined with a second colourisation network, which significantly increases the overall system complexity. This work reformulates the existing methodology introducing a novel end-to-end colourisation network that unifies the feature matching with the colourisation process. The proposed architecture integrates attention modules at different resolutions that learn how to perform the style transfer task in an unsupervised way towards decoding realistic colour predictions. Moreover, axial attention is proposed to simplify the attention operations and to obtain a fast but robust cost-effective architecture. Experimental validations demonstrate efficiency of the proposed methodology which generates high quality and visually appealing colourisation. Furthermore, the complexity of the proposed methodology is reduced compared to the state-of-the-art methods.
\end{abstract}

\blfootnote{The work described in this paper has been conducted within the project JOLT funded by the European Union’s Horizon 2020 research and innovation programme under the Marie Skłodowska Curie grant agreement No 765140.}

%%%%%%%%% BODY TEXT
\section{Introduction}

Colourisation refers to the process of adding colours to greyscale or other monochrome content such that the colourised results are perceptually meaningful and visually appealing. Digital colourisation  has become a classic task in computer vision, gaining significant importance in areas so diverse as broadcasting and film industries, restoration of legacy content or producer assistance. 

Although significant progress has been achieved, mapping colours from a grayscale input is a complex and ambiguous task due to the large degrees of freedom to arrive to a unique solution. In some cases, the semantics of the scene can help to infer priors of the colour distribution of the image, but in most cases the ambiguity in the decisions leads the system to make random choices, such as the colour of a car or a bird without further information. Thus, in order to overcome the ambiguity challenge, more conservative solutions propose the involvement of human interaction during the colour assignment process, introducing methodologies such as scribbled-based colourisation \cite{levin2004colorization, ironi2005colorization, huang2005adaptive, yatziv2006fast, qu2006manga, luan2007natural, zhang2017real, ci2018user, xu2009efficient} or exemplar-based colourisation \cite{bugeau2013variational, xu2020stylization, he2018deep, lu2020gray2colornet, reinhard2001color, chia2011semantic, welsh2002transferring, tai2005local, ironi2005colorization, pitie2007automated, morimoto2009automatic, gupta2012image, xiao2020example, zhang2019deep}. Specifically, colourisation by example can be automated by means of a retrieval system to select content related references, which can also be used as a recommender  for semi-automatic frameworks \cite{he2018deep}. However, existing methods are either highly sensitive to the selection of references (need of similar content, position and size of related objects) or extremely complex and time consuming. For instance, most exemplar-based approaches require a style transfer or similar method to compute the semantic correspondences between the target and the reference before starting the colourisation process. This fact usually increments the system complexity by requiring twofold pipelines with separate and even independent style transfer and colourisation systems. 

\begin{figure*}[t]
\begin{center}
   \includegraphics[width=1\linewidth]{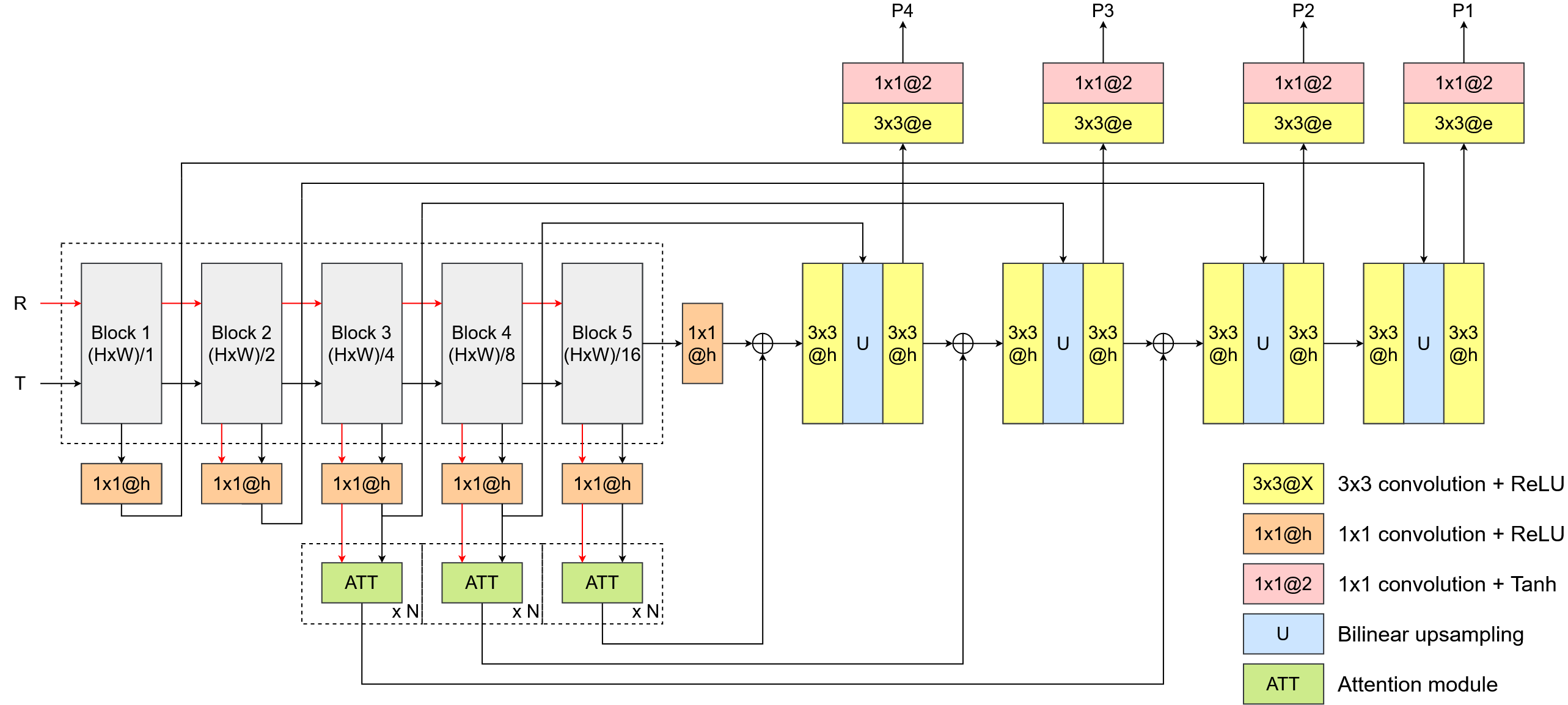}
\end{center}
   \caption{Proposed architecture for examplar-based image colourisation. $T$ is the black and white frame with luma component only, and $R$ is colour reference. Multi-scale outputs $P1$-$P4$ are used for training, where $P1$ are the colourised image components at targeted resolution.}
\label{fig:long}
\label{fig:att_architecture}
\end{figure*}

This work proposes a straightforward end-to-end solution which integrates attention modules that learn how to extract and transfer style features from the reference to the target in an unsupervised way during the colourisation process. Moreover, axial attention \cite{ho2019axial} is adopted to reduce the overall complexity and achieve a simple and fast architecture easily scalable to high resolution inputs. As shown in Figure \ref{fig:att_architecture}, the proposed architecture uses a pre-trained backbone to extract semantic and style features at different scales from the grayscale target and colour reference. Then, attention modules at different resolutions extract analogies between both feature sources and automatically yield output feature maps that fuse the style of the reference to the content of the target. Finally, a multi-scale pyramid decoder generates colour predictions at multiple resolutions, enabling the representation of higher-level semantics and robustness on the variance of scale and size of the local areas of content. The main advantage of such an end-to-end solution is that the attention modules learn how to perform style transfer based on the needs of the colourisation decoder in order to encourage high quality and realistic predictions, even if the reference significantly mismatches the target content. Moreover, it generalises the similarity computation of previous image analogy approaches in a way that does not constrain the similarity to a specific local patch search (attention modules can be interpreted as a set of long-term deformable kernels) and to specific similarity metrics. Finally, the proposed architecture introduces a novel design of the conventional transformer, enabling a modular combination of multi-head attention layers at different resolutions.

Overall, the contributions of this work are threefold:
\begin{itemize}
    \item A fast-end-to-end architecture for exemplar-based colourisation that improves existing methods while decreasing significantly the complexity and runtime.
    \item A multi-scale interpretation of the axial transformer for unsupervised style transfer and features analogy.
    \item A multi-loss training strategy that combines a multi-scale adversarial loss with conventional style transfer and exemplar-based colourisation losses.
\end{itemize}

%-------------------------------------------------------------------------
\section{Related work}

Modern digital colourisation algorithms can be roughly grouped into three main paradigms: automatic learning-based, scribble-based and exemplar-based colourisation.
Automatic learning-based methods can perform colourisation with end-to-end architectures which learn the direct mapping of every grayscale pixel to the colour space. Such approaches require large image datasets to train the network parameters without user intervention. However, in most cases they produce results which aren't colourful due to treating the colourisation process as a regression problem. As identified in the literature, well-designed loss functions such as adversarial loss \cite{isola2017image, blanch2019end}, classification loss \cite{larsson2016learning, vitoria2020chromagan} or perceptual loss \cite{wang2018high} or their combination with regularisation \cite{zhang2016colorful} is needed to better capture the colour distribution of the input content and enable more colourful results. A different approach is proposed in PixColor \cite{guadarrama2017pixcolor}, solving the automatic colourisation task as an autoregressive problem. Such methods predict the colour distribution of every pixel by conditioning to the grayscale input and the joint colour distribution of previous pixels. Similarly, ColTran \cite{kumar2021colorization} addresses the same methodology by using an axial transformer \cite{ho2019axial}. Autoregressive methods become impractical for colourisation due to the high dimensionality of the colour distribution and the related complexity of decoding high resolution images. For instance, even for modelling 8-bit RGB inputs only, the model needs to predict $256^3$ values. 

Scribble-based colourisation interactively propagates initial strokes or colour points annotated by the user to the whole grayscale image. An optimisation approach \cite{xu2009efficient} is proposed to propagate the user hints by using an adaptative clustering in the high dimensional affinity space. Alternatively, a Markov Random Field for propagating the scribbles \cite{levin2004colorization} is proposed under the rationale that adjacent pixels with similar intensity should have similar colours. Finally, a deep learning approach \cite{zhang2017real} fuses low-level cues along with high-level semantic information to propagate the user hints.

Exemplar-based colourisation uses a colour reference to condition the prediction process. An early approach proposed the matching of global colour statistics \cite{welsh2002transferring}, but yielded unsatisfactory results since it ignored local spatial information. More accurate approaches considered the extraction of correspondences at different levels, such as pixels \cite{liu2008intrinsic}, super-pixels \cite{gupta2012image, chia2011semantic}, segmented regions \cite{ironi2005colorization, tai2005local, bugeau2013variational} or deep features \cite{he2018deep, zhang2019deep}. Based on the extraction of deep image analogies from a pre-trained VGG-19 network \cite{simonyan2014very}, a deep learning framework uses previously computed similarity maps to perform exemplar-based colourisation \cite{he2018deep}. Such a method is posteriorly extended to video colourisation using  a temporal consistency loss to enforce temporal coherency \cite{he2018deep}. An alternative approach proposed the use of style transfer techniques based on AdaIN \cite{huang2017arbitrary} to generate an initial stylised version which is further refined with a colourisation network \cite{xu2020stylization}. Finally, a novel framework was proposed to fuse the semantic colours and global color distribution of the reference image towards the prediction of the final colour images \cite{lu2020gray2colornet}.

Finally, the architecture presented in this work adopts axial attention to reduce the complexity of the overall system. As introduced in the axial transformer \cite{ho2019axial}, attention is performed along a single axis, reducing the effective dimensionality of the attention maps and hence the complexity of the overall transformer. Such an approach managed to approximate conventional attention by focusing sequentially to each of the dimensions of the input tensor.  An application was proposed to perform panoptic segmentation \cite{wang2020axial}, integrating axial attention modules into a modified version of DeepLab \cite{chen2017deeplab}, and improving the original baseline.

%-------------------------------------------------------------------------
\section{Proposed method}

Aiming at exemplar-based colourisation, the goal of this method is to enable the colourisation of a grayscale target $T_L \in  \mathbb{R}^{1 \times H \times W}$ based on the colour of a reference $R_{Lab}  \in  \mathbb{R}^{3 \times H \times W}$, where $H \times W$ is the image dimension in pixels, represented in the \textit{CIE Lab} colour space \cite{connolly1997study}. Note that the target's $L$ index refers specifically to the luminance channel. To achieve this, an exemplar-based colourisation network is trained to model the mapping $\hat{T}_{ab} = F(T_L \mid R_{Lab})$ to the target \textit{ab} colour channels, conditioned to the reference $R_{Lab}$ channels. \textit{CIE Lab} colour space is chosen as it is designed to maintain perceptual uniformity and is more perceptually linear than other colour spaces \cite{connolly1997study}. This work assumes a normalised range of values between $[-1,1]$ for each of the Lab channels.

\subsection{Exemplar-based Colourisation Network}

As shown in Figure \ref{fig:att_architecture}, the proposed architecture is composed of four parts: the feature extractor backbone, the axial attention modules, the multi-scale pyramid decoder and the prediction heads.

First, both the target $T_L$ and the reference $R_{Lab}$ images are fed into a pre-trained feature extractor backbone to obtain $L$ multi-scale activated feature maps $F_T^l$, $F_R^l$ in an intermediate position of the $l$ convolutional block, where $l=\{1 \dots L\}$, and the last activated feature map only for the target input $F_T^B$, which is the output of the backbone. Note that the features have progressively coarser volumes with increasing levels. Without loss of generality, the experiments in this paper consider a VGG-19 network pre-trained on ImageNet \cite{deng2009imagenet}, extracting features $F_T^l$ and $F_R^l$ from the first Rectified Linear Unit (\textit{ReLU}) activation of every convolutional block (\textit{relu\{l\}\_1} from VGG-19), and $F_T^B$ which is the output of the encode (from \textit{relu\{5\}\_3} in VGG-19). Note that in order to feed $T_L$ into the pre-trained network, the luminace channel is triplicated to obtain a 3-dimensional input space. Then, all $F_T^l$, $F_R^l$ pairs and $F_T^B$ are projected onto a $h$-dimensional space by means of a $1 \times 1$ convolution plus \textit{ReLU} activation \cite{xu2015empirical}, to obtain $\hat{F}_T^{l}$, $\hat{F}_R^{l}$ and $\hat{F}_T^{B} \in \mathbb{R}^{h \times H_l \times W_l}$, respectively. 

Next, $N$ pairs ($\hat{F}_T^{l}$, $\hat{F}_R^{l}$), where $l=\{L-N+1 \dots L\}$, are fed into $N$ axial attention modules to compute a multi-head attention mask describing the deep correspondences between both sources. Then, the style of the reference source is transferred into the content of the target source by matrix multiplication of the attention mask with the reference source. Section \ref{sec:axial-attention} describes the axial attention module in depth and provides more information about the logic behind style transfer via attention. This process yields $N$ $h$-dimensional fused feature maps $\hat{F}_{TR}^l$.

After generating the multi-scale fused features, a multi-scale pyramid decoder composed on $L-1$ stacked decoders and prediction heads is employed to map $\hat{F}_T^{B}$ into $L-1$ colour predictions at different scales using the corresponding fused features $\hat{F}_{TR}^l$. Thus, starting with $O^{5}=\hat{F}_T^{B}$, each decoder $l=\{4, 3, 2, 1\}$ performs a five-fold operation: (1) adds $F_{TR}^l$ with the output of the previous decoder $O^{l-1}$, (2) applies a $3 \times 3$ convolution plus \textit{ReLU} activation, (3) upsamples the resultant feature map by a factor of $2$, (4) similar to the \textit{U-Net} architecture \cite{ronneberger2015u} concatenates the resultant upsampled map with the projected target feature map $F_T^l$ as skip connection and (5) refines the resultant map with another $3 \times 3$ convolution plus \textit{ReLU} activation which projects back the concatenated volume of $2h$ dimensions into the initial $h$ dimensions, yielding an output volume  $O^l$. 

Finally, the prediction heads map the decoded feature volumes $O^l$ into the output channels $\hat{T}_{ab}^l$. Each prediction head is composed of an $e$-dimensional $3 \times 3$ convolution plus \textit{ReLU} activation and $1 \times 1$ convolution plus hyperbolic tangent (\textit{Tanh}) activation to generate the $ab$ colour channels.

\subsection{Axial attention for unsupervised style transfer}
\label{sec:axial-attention}
\begin{figure}[t]
\begin{center}
   \includegraphics[width=1\linewidth]{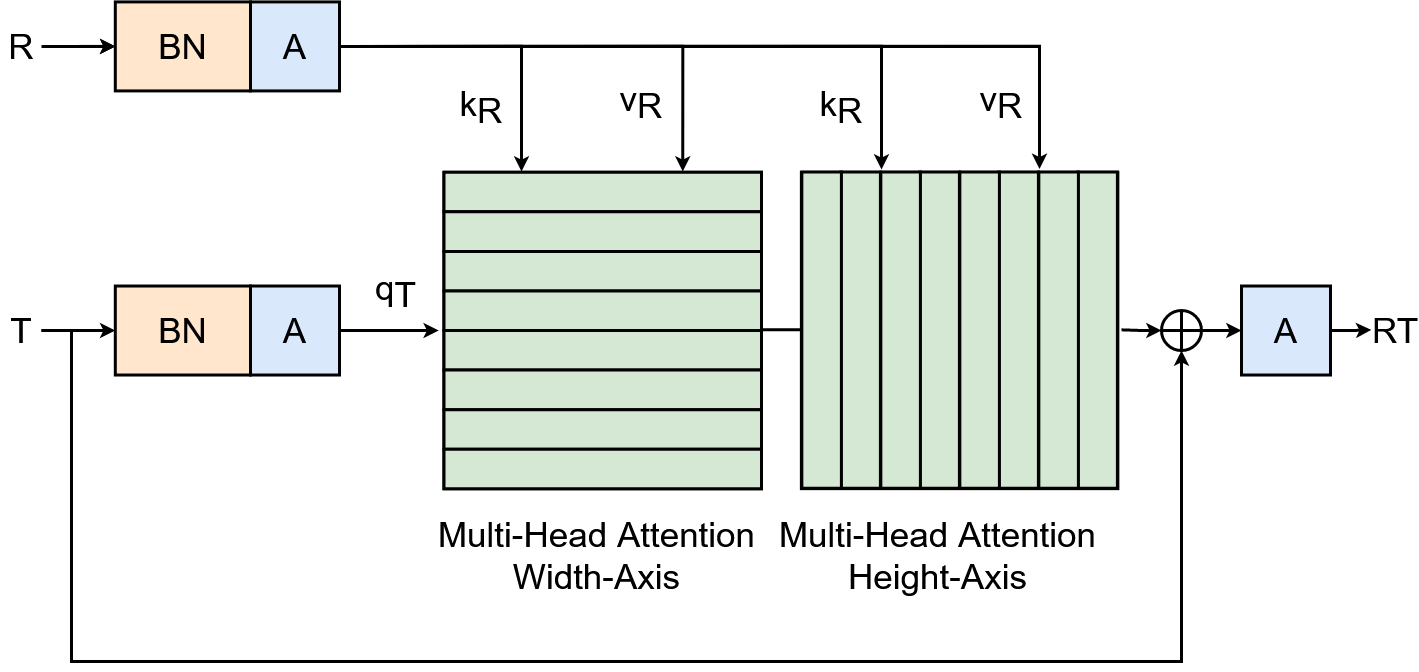}
\end{center}
   \caption{Axial attention module described in Section \ref{sec:axial-attention}. In the figure, BN means Batch Normalisation A activation.}
\label{fig:long}
\label{fig:att_module}
\end{figure}
Given two projected sources of features $\hat{F}_T^{l}$, $\hat{F}_R^{l}$  relative to the target and reference respectively, the goal of the axial attention module is to combine them in a way that the style codified in the reference features is transferred into similar content areas within the target features. 

Style transfer between two sources of features has been solved in many different ways, although in most cases only artistic style is targeted without contemplating the semantic analogies between both sources. Some strategies include the use of perceptual losses for training feed-forward networks for image transformation \cite{johnson2016perceptual}, in order to encourage the transformed images to produce similar features to the style reference when both are fed into a pre-trained loss network (e.g. VGG-16). A faster strategy is to use Adaptative Instance Normalisation (AdaIN) \cite{huang2017arbitrary} to align the mean and variance of the content features with those of the style features. Finally, another paradigm tackles deep image analogies for multi-scale visual attribute transfer \cite{liao2017visual}, but the analogy computation and transfer process are performed via a \textit{PatchMatch} algorithm \cite{barnes2009patchmatch}, which is computationally expensive.

This work proposes the use of attention to perform such processes faster and in an unsupervised way. In contrast with image analogy methods based on \textit{PatchMatch} algorithms, attention does not need to constrain to a specific local search technique (even if it can act as a set of long-term deformable kernels) nor the similarity metric (e.g. correlation loss, cosine similarity) since the module learns it automatically. Attention was introduced to tackle the problem of long-range interactions in sequence modelling \cite{vaswani2017attention, lin2017structured, parikh2016decomposable, cheng2016long}. However, attention modules have been employed recently to improve computer vision tasks such as object detention \cite{carion2020end} or image classification \cite{dosovitskiy2020image}, by providing contextual information from other sources of information. Following the same rationale, an attention mechanism can solve the semantic analogy problem in style transfer by focusing on the most relevant areas of the style source when decoding each voxel in the content source.

Following the original definition of stand-alone attention \cite{vaswani2017attention, wang2018non, zhang2019self}, given a projected target and reference feature maps $\hat{F}_T^{l}$, $\hat{F}_R^{l}$, the fused feature map at position $o=(i,j)$, $\hat{F}_{TR(o)}^l$ is computed as follows:
\begin{equation}
\hat{F}_{TR(o)} = \sum_{p\in \EuScript{N}^l} \text{\textit{softmax}}_{p}(\hat{q}_{T(o)}^{l \intercal} \hat{k}_{R(p)}^{l})\hat{v}_{R(p)}^{l},\label{eq1}
\end{equation}
\noindent where $\EuScript{N}^l \in \mathbb{R}^{H_l \times W_l}$ is the whole 2D location lattice. Furthermore, the queries to the target source  ${\hat{q}_{T(o)}^{l}}=W_q {\hat{F}_{T(o)}^{l}}$ and the keys and values from the reference source ${\hat{k}_{R(o)}^{l}}=W_k {\hat{F}_{R(o)}^{l}}$, ${\hat{v}_{R(o)}^{l}}=W_v {\hat{F}_{R(o)}^{l}}$ are all linear projections of the target and reference projected sources $\hat{F}_{T(o)}^{l}$ and  ${\hat{F}_{R(o)}^{l}}$, respectively, $\forall o \in \EuScript{N}$, where $W_q, W_k, W_v \in \mathbb{R}^{h \times h}$ are all the learnable parameters. The $\text{\textit{softmax}}_p$ denotes a softmax operation applied to all possible $p$ positions within the 2D lattice $\EuScript{N}^l$.

Next, a position-sensitive learned positional encoding \cite{cordonnier2019relationship, ramachandran2019stand, wang2020axial} is adopted to encourage the attention modules to model dynamic prior of where to look at in the receptive field of the reference source ($m \times m$ region within $\EuScript{N}^l$). Positional encoding has proven to be beneficial in computer vision tasks to exploit spatial information and capture shapes and structures within the sources of input features. Therefore, as in \cite{wang2020axial}, a key, query and value dependent positional encoding are applied to Equation \ref{eq1} as follows:
\begin{equation}
\begin{split}
    \hat{F}_{TR(o)} = \sum_{p\in \EuScript{N}^l_{m \times m (o)}} & \text{\textit{softmax}}_p({\hat{q}_{T(o)}^{l \intercal}} {\hat{k}_{R(p)}^{l}} +  {\hat{q}_{T(o)}^{l \intercal}} r^q_{(p-o)} + \\ 
    & + {\hat{k}_{T(p)}^{l \intercal}} r^k_{(p-o)})({\hat{v}_{R(p)}^{l}} + r^v_{(p-o)}),\label{eq2}
\end{split}
\end{equation}
\noindent where $\EuScript{N}^l_{m \times m (o)}$ is the local $m \times m$ local region centred around location $o=(i,j)$, and $r_{(p-o)}^q, \ r_{(p-o)}^k and \ r_{(p-o)}^v$ the learned relative positional encoding for queries, keys and values, respectively. The inner products ${\hat{q}_{T(o)}^{l \intercal}} r^q_{(p-o)}$ and ${\hat{k}_{T(p)}^{l \intercal}} r^k_{(p-o)}$ measure the compatibilities from location $p$ to $o$ within the queries and keys space, and $r_{(p-o)}^v$ guides the output $\hat{F}_{TR(o)}$ to retrieve content within the values space.

Finally, axial attention \cite{ho2019axial} is adopted to reduce the complexity of the original formulation $\mathcal{O}(H_lW_lm^2)$ to $\mathcal{O}(H_lW_lm)$ by computing the attention operations along a $1$-dimensional axial lattice $1 \times m$, instead of across the whole $\{m \times m\}$ space. Following the formulation as in stand-alone axial-DeepLab \cite{wang2020axial}, the global attention operation is simplified by defining an axial-attention layer that propagates the information along the width-axis followed by another one along the height-axis. In this work, we set a span $m=\{H_l,W_l\}$ equal to the input image resolution ($\mathcal{O}(H_lW_lm$)), but such values can be reduced for high resolution inputs. Finally, multi-head attention can be performed by applying $N$ single axial attention heads with head-dependent projections $W_q^n, \ W_k^n, \ W_v^n$, posteriorly concatenating the results of each head and projecting the final output maps by means an output $1 \times 1$ convolution.

As shown in Figure \ref{fig:att_module}, a succession of multi-head \textit{weight-height} axial attention layers are integrated to design the axial attention module for unsupervised style transfer. Given $F_T^{lh}$, $F_R^{lh}$ inputs, such module performs a three-fold operation: (1) normalise the target and reference projected sources by means of batch normalisation plus \textit{ReLU} activation, (2) fuse the normalised sources by means of the multi-head \textit{weight-height} axial attention layers, and (3) add resulting features to the target source identity $F_T^{lh}$ plus activate the output with a \textit{ReLU} activation.

\begin{figure*}[t]
\begin{center}
   \begin{tabular}{cccccc}
\includegraphics[width=0.13\linewidth]{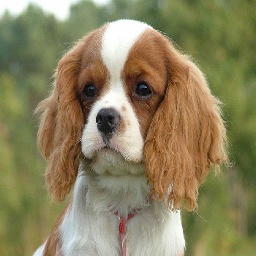} & 
\includegraphics[width=0.13\linewidth]{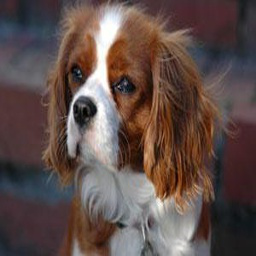} & 
\includegraphics[width=0.13\linewidth]{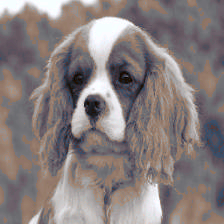} & 
\includegraphics[width=0.13\linewidth]{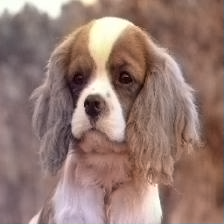} & 
\includegraphics[width=0.13\linewidth]{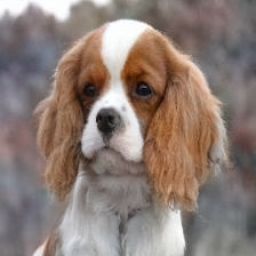} & 
\includegraphics[width=0.13\linewidth]{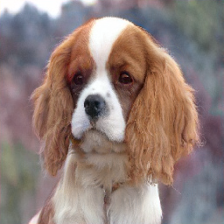}  \\
\includegraphics[width=0.13\linewidth]{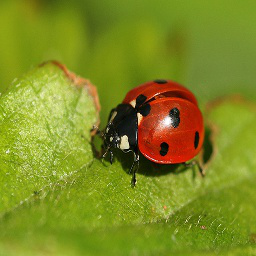} & 
\includegraphics[width=0.13\linewidth]{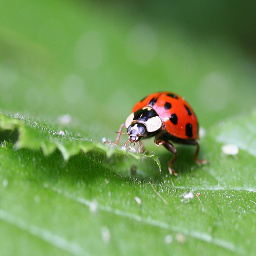} & 
\includegraphics[width=0.13\linewidth]{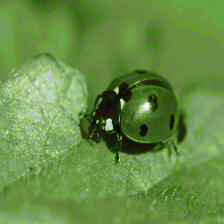} & 
\includegraphics[width=0.13\linewidth]{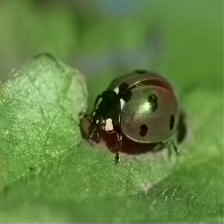} & 
\includegraphics[width=0.13\linewidth]{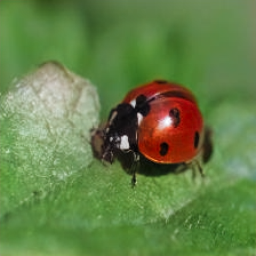} & 
\includegraphics[width=0.13\linewidth]{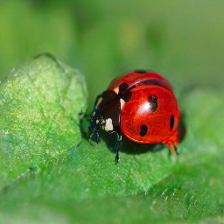}  \\
\includegraphics[width=0.13\linewidth]{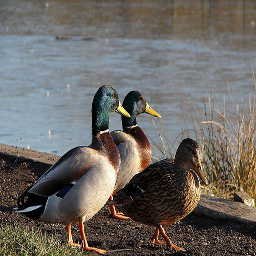} & 
\includegraphics[width=0.13\linewidth]{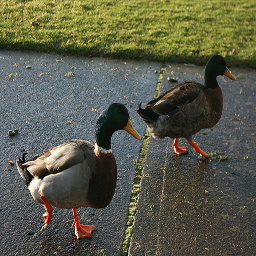} & 
\includegraphics[width=0.13\linewidth]{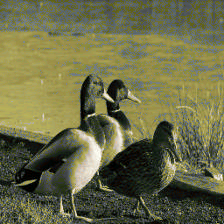} & 
\includegraphics[width=0.13\linewidth]{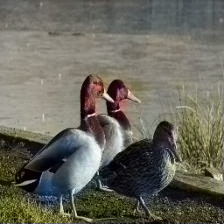} & 
\includegraphics[width=0.13\linewidth]{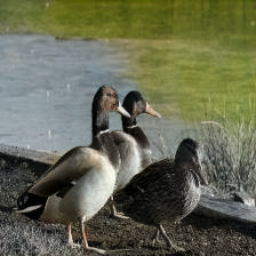} & 
\includegraphics[width=0.13\linewidth]{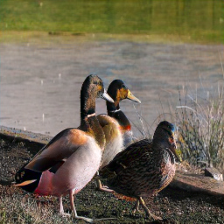}  \\
\includegraphics[width=0.13\linewidth]{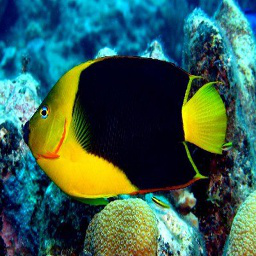} & 
\includegraphics[width=0.13\linewidth]{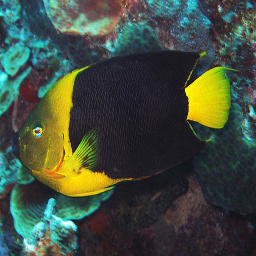} & 
\includegraphics[width=0.13\linewidth]{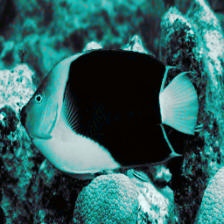} & 
\includegraphics[width=0.13\linewidth]{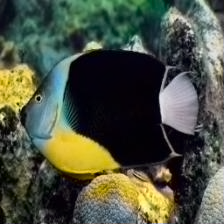} & 
\includegraphics[width=0.13\linewidth]{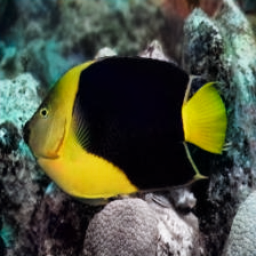} & 
\includegraphics[width=0.13\linewidth]{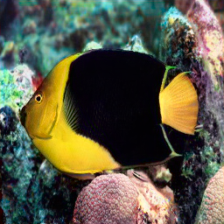}  \\

Target & 
Reference & 
Welsh et al. \cite{welsh2002transferring} & 
Xiao et al. \cite{xiao2020example} & 
Zhang et al. \cite{zhang2019deep} & 
Ours \\
\end{tabular}
\end{center}
   \caption{Qualitative comparison of the existing and the proposed exemplar-based colourisation methods.}
\label{fig:visual_comparison}
\end{figure*}
\subsection{Training losses}

Usually, the objective of colourisation is to encourage that the predicted $\hat{T}_{ab}$ colour channels are as close as possible to the ground truth $T_{ab}$ in the original training dataset. However, this objective does not apply in exemplar-based colourisation, where $\hat{T}_{ab}$ should be customized by the colour reference $R_{Lab}$ while preserving the content of the grayscale target $T_{L}$. Therefore, the definition of the training strategy is not straightforward, as penalising $\hat{T}_{ab}$ and $T_{ab}$ is not accurate. Then, the objective is to enable the reliable transfer of reference colours to the target content towards obtaining a colour prediction faithful to the reference. This work takes advantage of the pyramidal decoder to combine state-of-the-art exemplar-based losses with an adversarial training at multiple resolutions. Hence, a multi-loss training strategy is proposed to combine a \textit{smooth-}$L_1$ loss, a colour histogram loss and a total variance regularisation, as in \cite{morimoto2009automatic}, with a multi-scale adversarial loss by means of multiple patch-based discriminators \cite{isola2017image}. In order to handle multi-scale losses, average pooling with a factor of $2$ is applied to both target and reference to successively generate the multi-scale ground truth $T_{ab}^l$ and $R_{ab}^l$.

\textbf{\textit{Smooth-}$L_1$ loss}. In order to induce dataset priors in cases when the content of reference highly mismatches with the target, a pixel loss $L_{pixel}^l$ based on Huber loss \cite{huber1992robust} (also known as \textit{smooth-}$L_1$) is proposed to encourage realistic predictions. Huber loss is widely used in colourisation as a substitute of the standard $L_1$ loss in order to avoid the averaging solution in the ambiguous colourisation problem \cite{zhang2016colorful}. As spotted in the Fast R-CNN paper \cite{ren2015faster}, \textit{smooth-}$L_1$ is less sensitive to outliers than the $L_2$ loss and in some cases prevents exploding gradients.

\textbf{Colour histogram loss}. As proposed in \cite{lu2020gray2colornet}, in order to fully capture the global colour distribution of the reference image and penalise the differences with the predicted colour distribution, a colour histogram loss is considered. The computation of the colour histogram is not a differentiable process which can be integrated within the training loop. To avoid this problem, the aforementioned exemplar-based colourisation approach \cite{lu2020gray2colornet} approximates the colour histograms $\mathcal{H}^l_R$ and $\mathcal{H}_{\hat{T}}^l$, corresponding to the reference and predicted images respectively, by means of a function similar to a bilinear interpolation. Then, the histogram loss $L_{hist}^l$ is defined as a symmetric $\chi^2$ distance \cite{puzicha1997non} as follows:
\begin{equation}
L_{hist}^l = 2\sum_{q=1}^Q {\frac{\left(\mathcal{H}_{\hat{T}}^l(q) - \mathcal{H}^l_R(q) \right)^2}{\mathcal{H}_{\hat{T}}^l(q) + \mathcal{H}^l_R(q) + \epsilon}}, \label{eq6}
\end{equation}
\noindent where $\epsilon$ prevents infinity overflows and $Q$ is the number of histogram bins. In this work, $\epsilon = 10^{-5}$ and $Q=441$.

\textbf{Total variance regularisation}. As widely used in style transfer literature \cite{johnson2016perceptual}, a total variance loss $L_{TV}^l$ is proposed in order to encourage low variance along neighbouring pixels of the predicted colour channels $\hat{T}_{ab}^l$.

\textbf{Adversarial loss}. Although the histogram loss encourages the predictions to contain reference colours, it does not consider spatial information nor discriminate between how realistic different object instances are colourised. With the aim to guide the previous losses towards realistic decisions, an adversarial strategy based on \textit{LS-GAN} \cite{mao2017least} is proposed to derive the scale-based generator loss $L_G^l$ and discriminator loss $L_D^l$, using the ground truth colour targets  $T^l$ as real sources and a patch-based discriminator D (same as used in \cite{isola2017image}). Note that within the GAN framework, the proposed exemplar-based colourisation network would be the generator. Then, the total discriminator loss $L_D$ is computed by adding the $L$ individual multi-scale losses: $L_D = \sum_{l=1}^L L_{D}^l$

Finally, the total multi-scale loss $L_{total}$ is computed as:
\begin{equation}
\begin{split}
    L_{total} = & \sum_{l=1}^L (\lambda_{pixel}L_{pixel}^l + \lambda_{hist}L_{hist}^l + \\
    & + \lambda_{TV}L_{TV}^l + \lambda_{G}L_{G}^l), \label{eq10}
\end{split}
\end{equation}
\noindent where $\lambda_{pixel}, \ \lambda_{hist}, \ \lambda_{TV}$ and $\lambda_{G}$ are the multi-loss weights which specify the contribution of each individual loss.

%-------------------------------------------------------------------------
\section{Experiments}

\subsection{Training settings}
\label{training-settings}
A training dataset based on \textit{ImageNet} \cite{deng2009imagenet} is generated by sampling $225,000$ images from the $750$ most popular categories ($300$ images per class), which include: animals, plants, people, scenery, food, transportation and artifacts. Pairs of target-reference images are randomly generated based on the correspondence recommendation pipeline proposed in \cite{he2018deep}. First, a top-5 global ranking is created by minimising the $L_2$ distance between the features of the target and the rest of the of the same class, extracted at the first fully connected layer of a pre-trained VGG-19 with ImageNet and projected into $128$ dimensions via PCA transformation \cite{babenko2014neural}. Next, following the process in \cite{he2018deep}, the global ranking is refined by a local search selecting the most similar image by means of a patch-based similarity. The top-1 reference is selected by minimising the cosine distance between $16 \times 16$ patches corresponding to the most similar position-wise feature vector at the \textit{relu\{4\}\_3} space of the same pre-trained VGG-19, from both target and reference candidate. Finally, pairs of target-reference images are randomly sampled on-the-fly during training by using a weighted uniform distribution of $3$ categories, with a weight $\alpha_c$: top-1 reference ($\alpha_1 = 0.6$), random choice among the top-5 candidates ($\alpha_2 = 0.3$) and random choice among the rest of images of the same class ($\alpha_3 = 0.1$). Testing data is generated in a similar way, sampling $45,000$ pairs of target-reference images from the training subset (different targets than in training) at the same categories ($60$ images per class). All  images are resized to $224 \times 224$ pixels, converted to the CIE Lab colour space and normalised into the range $[-1,1]$ for each channel.

All the experiments use multi-head attention layers of $8$ heads, a hidden dimension $h=256$ and a prediction head dimension $e=64$. As shown in the Figure \ref{fig:att_architecture}, a backbone with $5$ convolutional blocks is used, starting the decoding process from a resolution of $(H \times W)/16$ pixels and decoding $4$ different multi-scale predictions. Although several ablations are performed, the best trade-off between complexity and performance is achieved by applying the attention modules from the block $3$. All models are trained around $30$ epochs using an Adam optimiser \cite{kingma2014adam} with a learning rate of $10^{-5}$. The multi-loss weights $\lambda_{pixel}=100, \ \lambda_{hist}=2, \ \lambda_{TV}=50$ and $\lambda_{G}=1$ are used for all the experiments. Finally, all models are implemented in Pytorch 1.7.0 \cite{paszke2019pytorch} and trained with a single GPU using a batch size of around $4-12$ samples.

\subsection{Comparison with colourisation methods}
In order to compare our approach with existing exemplar-based colourisation methods \cite{zhang2019deep, xiao2020example, welsh2002transferring}, a test dataset is collected by randomly sampling $5,000$ target-reference pairs from the validation set defined in Section \ref{training-settings}. To provide a fair comparison, all results are obtained by running the original publicly available codes and models provided by the authors. 

A qualitative comparison for a selection of representative cases is shown in Figure \ref{fig:visual_comparison}. From this comparison, our method along with Zhang et al. \cite{zhang2019deep} produce the most visual appealing results, being able to transfer effectively the colours from the reference. Both methods show that image analogy methodology better captures local information from semantically related objects and leads to more precise colour predictions. On the contrary, the methods from Welsh et al. \cite{welsh2002transferring} and Xiao et al. \cite{xiao2020example}, based on global histogram estimation, fail to detect precise patterns and only map overall tones from the reference. The proposed multi-loss strategy, incorporating histogram and adversarial loss at different resolutions, enables more colorful and saturated results. However, unlike the conservative colourisation of \cite{zhang2019deep}, the instability of the adversarial training can lead to some colour noise, as can be seen in the 4th row of Figure \ref{fig:visual_comparison}. A better control of the adversarial loss could boost our method's performance, reaching the stability of \cite{zhang2019deep} while producing more colourful and visually appealing predictions.
\begin{figure}[t]
\begin{center}
   \includegraphics[width=0.9\linewidth]{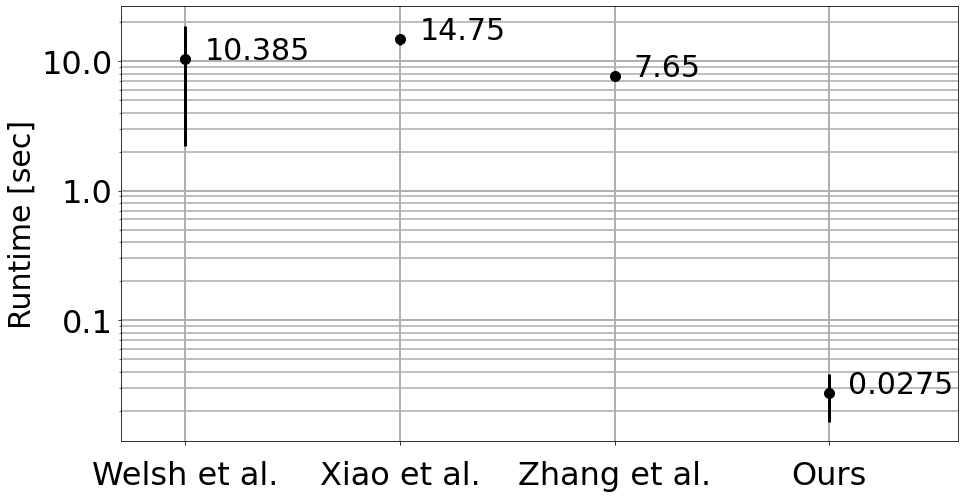}
\end{center}
   \caption{Runtime comparison in seconds.}
\label{fig:long}
\label{fig:runtime}
\end{figure}
\begin{table}
\begin{center}
\begin{tabular}{|r|c|c|c|c|}
\hline 
Method & HIS & SSIM & \begin{tabular}[c]{@{}c@{}}Top-1\\ Acc\end{tabular} & \begin{tabular}[c]{@{}c@{}}Top-5\\ Acc\end{tabular} \\
\hline\hline
Welsh et al. \cite{welsh2002transferring}  & 0.55  & 0.78 & 50.3\% & 74.1\%  \\
Xiao et al. \cite{xiao2020example}      & 0.59 & 0.84 & 54.8\% & 79.2\% \\
Zhang et al. \cite{zhang2019deep}      & 0.66 & \textbf{0.88} & 65.6\% & 84.8\% \\
\hline
\textit{Ours axial att.}         & 0.72 & 0.87 & 68.1\% & 89.1\% \\
\textbf{Ours standard att}.      & \textbf{0.74} & \textbf{0.88} & \textbf{69.7\%} & \textbf{90.1\%} \\
Ours single module      & 0.68 & \textbf{0.88} & 67.6\% & 88.9\% \\
Ours w/o adv. loss      & 0.70 & \textbf{0.88} & 67.5\% & 89.2\% \\
Ours w/o pix. loss      & 0.68 & 0.86 & 67.5\% & 86.7\% \\
Ours w/o hist. loss     & 0.54 & \textbf{0.88} & 65.4\% & 89.2\% \\
\hline
\end{tabular}
\end{center}
\caption{Quantitative comparison of the state-of-the art methods with the proposed method in different settings. Note that standard attention is only used in the ablation study, the rest of our combinations use axial attention.}
\label{quantitative-evaluation}
\end{table}
\begin{figure*}[t]
\begin{center}
   \begin{tabular}{cccccc}
\includegraphics[width=0.13\linewidth]{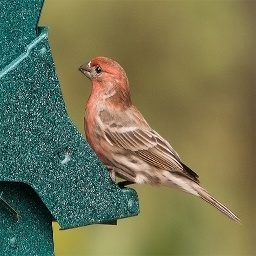} & 
\includegraphics[width=0.13\linewidth]{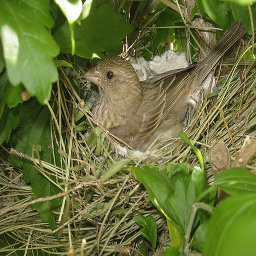} & 
\includegraphics[width=0.13\linewidth]{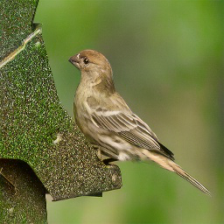} & 
\includegraphics[width=0.13\linewidth]{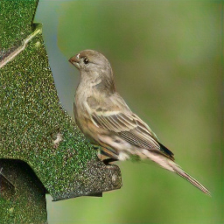} & 
\includegraphics[width=0.13\linewidth]{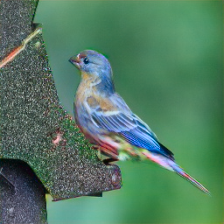} &  
\includegraphics[width=0.13\linewidth]{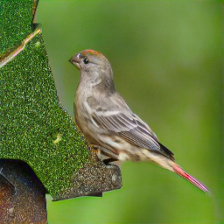} \\
Target & 
Reference & 
Ours & 
w/o adv. loss & 
w/o hist. loss &
w/o pix. loss\\
\end{tabular}
\end{center}
   \caption{Visual comparison of each individual training loss contribution.}
\label{fig:losses-ablation}
\end{figure*}

\begin{figure*}[t]
\begin{center}
   \begin{tabular}{ccccc}
\includegraphics[width=0.13\linewidth]{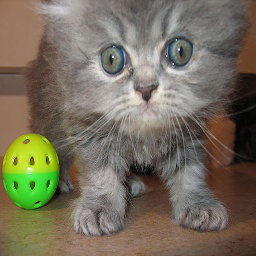} & 
\includegraphics[width=0.13\linewidth]{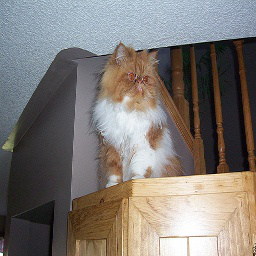} & 
\includegraphics[width=0.13\linewidth]{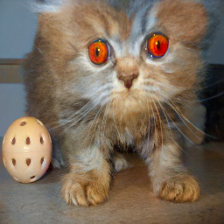} & 
\includegraphics[width=0.13\linewidth]{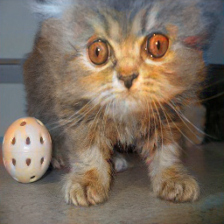} & 
\includegraphics[width=0.13\linewidth]{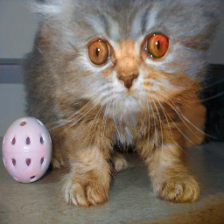}  \\
Target & 
Reference & 
Standard & 
Axial multiple & 
Axial single \\
\end{tabular}
\end{center}
   \caption{Visual comparison of the attention module configurations, using standard attention or axial attention one time or two times. }
\label{fig:att-ablation}
\end{figure*}
Moreover, a quantitative comparison is shown in Table \ref{quantitative-evaluation}, using three different metrics: Histogram Intersection Similarity (HIS) \cite{isola2017image} relative to the reference image, Structural Similarity Index Measure (SSIM) relative to the target ground truth image and classification accuracy. First, HIS score measures the averaged colour histogram intersection between the reference and predicted images. As shown in the results, our method along with \cite{zhang2019deep}, which are both based on semantic-related analogies, achieve higher HIS scores suggesting a better transfer of the reference colours. On the contrary, the methods in \cite{xiao2020example} and \cite{welsh2002transferring}, based on global histogram estimation, slightly lower the HIS score due to the averaged colourisation in ambiguous cases where the target and reference objects are not recognised. SSIM score is used to estimate structural similarity of each method. As can be observed, the methods achieving a more precise colourisation obtain higher SSIM score. The method in \cite{zhang2019deep} achieves the same score as ours, suggesting that more stable predictions help to better retain the structural information of the target image. Finally, our method outperforms all other methods on image recognition accuracy when the colour predictions are fed into a VGG-16 pre-trained on ImageNet. The obtained results indicate that the proposed method overall outperforms previous methods, which is also reflected by the classification performance.

\begin{table}
\begin{center}
\begin{tabular}{|r|c|}
\hline
Method & Naturalness (\%) \\
\hline\hline
Real images & $85.51\%$ \\
\textbf{Ours} & $\textbf{61.30\%}$ \\
Zhang et al. \cite{zhang2019deep} & $60.04\%$ \\
Xiao et al. \cite{xiao2020example} & $46.10\%$ \\
Welsh et al. \cite{welsh2002transferring} & $36.10\%$ \\
\hline
\end{tabular}
\end{center}
\caption{Perceptual test results. The values show the percentage of images selected as genuine (natural) for each of the methods.}
\label{perceptual-test}
\end{table}

In addition to qualitative and quantitative comparisons, a perceptual test is performed to validate overall results and to detect possible failure cases. A total of $100$ target-reference pairs are randomly sampled from the test dataset and colourised using our method and the three state-of-the-art methods \cite{zhang2019deep, xiao2020example, welsh2002transferring}. Therefore, $500$ images are generated, including $100$ original images and $400$ images which the colour is predicted. Each individual test session randomly selects $25$ images and shows them one by one to viewers, which included participants with technical and non-technical backgrounds. Then, each participant has to annotate if the colours in each image appear to be genuine (natural) or not. The study was performed $190$ times, generating a total of $4,750$ annotations. As shown in Table \ref{perceptual-test} which shows the percentage of the annotations that evaluated colours as genuine with reference to the total number of all the annotations for the specific method, our approach ($61.30\%$) slightly outperforms the method in \cite{zhang2019deep} ($60.04\%$). As discussed in the visual comparison, the potential production of colour noise might have lowered the performance of our method. In contrast, the stability of \cite{zhang2019deep} enabled a considerably high rate despite its conservative colourisation. Finally, the methods in \cite{xiao2020example} ($46.10\%$) and \cite{welsh2002transferring} ($36.10\%$) achieve the lowest results.

Finally, the runtime is also compared to highlight the efficiency of the proposed end-to-end architecture. All the results are obtained using the implementation provided by the authors. Runtime values are obtained on a machine with 3.60GHz Intel Xeon Gold 5122 CPU and a single NVIDIA GeForce RTX 2080 Ti GPU. As shown in Figure \ref{fig:runtime}, among neural network based methods, the pyramid structure in Xiao et al. \cite{xiao2020example} costs most of the time. The method from Zhang et al. \cite{zhang2019deep} slightly reduces runtime but the Patch Match search used in Deep Image Analogy \cite{liao2017visual} consumes a lot of time. On contrast, our end-to-end approach significantly reduces complexity achieving runtimes of 20 ms per image.

\subsection{Ablation study}
Several experiments are performed with the aim to evaluate the effects of the different architectural choices and training hyperparameters. The ablation study includes the analysis of the attention module, comparing the performance obtained with the standard attention operation and the proposed axial attention simplifications in Section \ref{sec:axial-attention}. Moreover, the number of attention operations at each scale is also evaluated. Finally, the contribution of each of the training losses is validated by removing them separately from the total multi-loss function and study their effects on the final predictions. As discussed in Section \ref{sec:axial-attention}, axial attention is adopted to reduce the complexity of the original attention formulation $\mathcal{O}(H_lW_lm^2)$ to $\mathcal{O}(H_lW_lm)$ by computing the attention operations along a single $1$-dimensional axis, instead of across the whole $\{m \times m\}$ space. Although axial attention is applied to both the horizontal and vertical axis to approximate the standard performance, a significant loss is identified in Table \ref{quantitative-evaluation}. A visual comparison is shown in Figure \ref{fig:att-ablation}, where standard attention yields to more precise results, being able to capture longer-term relationships. In order to refine the axial approximation and being able to derive more complex relationships, the attention module is applied 2 consecutive times. As shown, such approach outperforms the single configuration in both quantitative and qualitative evaluations. Finally, the individual contribution of each training loss is evaluated by removing them one by one from the multi-loss configuration. As shown in Table \ref{quantitative-evaluation} and Figure \ref{fig:losses-ablation}, a major drop in HIS score is identified in the absence of the histogram loss, indicating its importance to guide the learning process towards an effective transfer of reference colours. The absence of the adversarial loss also lowers the performance, dropping by 0.2 the HIS score and 0.6\% the Top-1 accuracy. However, a higher effect is shown in the visual comparison, where a clear loss of both colourfulness and naturalness can be observed.

%-------------------------------------------------------------------------
\section{Conclusions}
%-------------------------------------------------------------------------
This paper introduces a novel architecture for exemplar-based colourisation. The proposed model integrates attention modules at different resolutions that learn how to perform style transfer in an unsupervised way towards decoding realistic colour predictions. Such methodology significantly simplifies previous exemplar-based approaches, unifying the feature matching with the colourisation process and therefore achieving a fast end-to-end colourisation. Moreover, in order to further reduce the model complexity, axial attention is proposed to simplify the standard attention operations and hence reduce the computation intensity. The proposed method outperforms state-of-the-art methods in both visual quality and complexity, and significantly reduces the runtime.

{\small
\bibliographystyle{ieee_fullname}
\bibliography{egbib}

\begin{thebibliography}{10}\itemsep=-1pt

\bibitem{babenko2014neural}
Artem Babenko, Anton Slesarev, Alexandr Chigorin, and Victor Lempitsky.
\newblock Neural codes for image retrieval.
\newblock In {\em European conference on computer vision}, pages 584--599.
  Springer, 2014.

\bibitem{barnes2009patchmatch}
Connelly Barnes, Eli Shechtman, Adam Finkelstein, and Dan~B Goldman.
\newblock Patchmatch: A randomized correspondence algorithm for structural
  image editing.
\newblock {\em ACM Trans. Graph.}, 28(3):24, 2009.

\bibitem{blanch2019end}
Marc~Gorriz Blanch, Marta Mrak, Alan~F Smeaton, and Noel~E O'Connor.
\newblock End-to-end conditional gan-based architectures for image
  colourisation.
\newblock In {\em 2019 IEEE 21st International Workshop on Multimedia Signal
  Processing (MMSP)}, pages 1--6. IEEE, 2019.

\bibitem{bugeau2013variational}
Aur{\'e}lie Bugeau, Vinh-Thong Ta, and Nicolas Papadakis.
\newblock Variational exemplar-based image colorization.
\newblock {\em IEEE Transactions on Image Processing}, 23(1):298--307, 2013.

\bibitem{carion2020end}
Nicolas Carion, Francisco Massa, Gabriel Synnaeve, Nicolas Usunier, Alexander
  Kirillov, and Sergey Zagoruyko.
\newblock End-to-end object detection with transformers.
\newblock In {\em European Conference on Computer Vision}, pages 213--229.
  Springer, 2020.

\bibitem{chen2017deeplab}
Liang-Chieh Chen, George Papandreou, Iasonas Kokkinos, Kevin Murphy, and Alan~L
  Yuille.
\newblock Deeplab: Semantic image segmentation with deep convolutional nets,
  atrous convolution, and fully connected crfs.
\newblock {\em IEEE transactions on pattern analysis and machine intelligence},
  40(4):834--848, 2017.

\bibitem{cheng2016long}
Jianpeng Cheng, Li Dong, and Mirella Lapata.
\newblock Long short-term memory-networks for machine reading.
\newblock {\em arXiv preprint arXiv:1601.06733}, 2016.

\bibitem{chia2011semantic}
Alex Yong-Sang Chia, Shaojie Zhuo, Raj~Kumar Gupta, Yu-Wing Tai, Siu-Yeung Cho,
  Ping Tan, and Stephen Lin.
\newblock Semantic colorization with internet images.
\newblock {\em ACM Transactions on Graphics (TOG)}, 30(6):1--8, 2011.

\bibitem{ci2018user}
Yuanzheng Ci, Xinzhu Ma, Zhihui Wang, Haojie Li, and Zhongxuan Luo.
\newblock User-guided deep anime line art colorization with conditional
  adversarial networks.
\newblock In {\em Proceedings of the 26th ACM international conference on
  Multimedia}, pages 1536--1544, 2018.

\bibitem{connolly1997study}
Christine Connolly and T Fleiss.
\newblock A study of efficiency and accuracy in the transformation from rgb to
  cielab color space.
\newblock {\em IEEE transactions on image processing}, 6(7):1046--1048, 1997.

\bibitem{cordonnier2019relationship}
Jean-Baptiste Cordonnier, Andreas Loukas, and Martin Jaggi.
\newblock On the relationship between self-attention and convolutional layers.
\newblock {\em arXiv preprint arXiv:1911.03584}, 2019.

\bibitem{deng2009imagenet}
Jia Deng, Wei Dong, Richard Socher, Li-Jia Li, Kai Li, and Li Fei-Fei.
\newblock Imagenet: A large-scale hierarchical image database.
\newblock In {\em 2009 IEEE conference on computer vision and pattern
  recognition}, pages 248--255. Ieee, 2009.

\bibitem{dosovitskiy2020image}
Alexey Dosovitskiy, Lucas Beyer, Alexander Kolesnikov, Dirk Weissenborn,
  Xiaohua Zhai, Thomas Unterthiner, Mostafa Dehghani, Matthias Minderer, Georg
  Heigold, Sylvain Gelly, et~al.
\newblock An image is worth 16x16 words: Transformers for image recognition at
  scale.
\newblock {\em arXiv preprint arXiv:2010.11929}, 2020.

\bibitem{guadarrama2017pixcolor}
Sergio Guadarrama, Ryan Dahl, David Bieber, Mohammad Norouzi, Jonathon Shlens,
  and Kevin Murphy.
\newblock Pixcolor: Pixel recursive colorization.
\newblock {\em arXiv preprint arXiv:1705.07208}, 2017.

\bibitem{gupta2012image}
Raj~Kumar Gupta, Alex Yong-Sang Chia, Deepu Rajan, Ee~Sin Ng, and Huang
  Zhiyong.
\newblock Image colorization using similar images.
\newblock In {\em Proceedings of the 20th ACM international conference on
  Multimedia}, pages 369--378, 2012.

\bibitem{he2018deep}
Mingming He, Dongdong Chen, Jing Liao, Pedro~V Sander, and Lu Yuan.
\newblock Deep exemplar-based colorization.
\newblock {\em ACM Transactions on Graphics (TOG)}, 37(4):1--16, 2018.

\bibitem{ho2019axial}
Jonathan Ho, Nal Kalchbrenner, Dirk Weissenborn, and Tim Salimans.
\newblock Axial attention in multidimensional transformers.
\newblock {\em arXiv preprint arXiv:1912.12180}, 2019.

\bibitem{huang2017arbitrary}
Xun Huang and Serge Belongie.
\newblock Arbitrary style transfer in real-time with adaptive instance
  normalization.
\newblock In {\em Proceedings of the IEEE International Conference on Computer
  Vision}, pages 1501--1510, 2017.

\bibitem{huang2005adaptive}
Yi-Chin Huang, Yi-Shin Tung, Jun-Cheng Chen, Sung-Wen Wang, and Ja-Ling Wu.
\newblock An adaptive edge detection based colorization algorithm and its
  applications.
\newblock In {\em Proceedings of the 13th annual ACM international conference
  on Multimedia}, pages 351--354, 2005.

\bibitem{huber1992robust}
Peter~J Huber.
\newblock Robust estimation of a location parameter.
\newblock In {\em Breakthroughs in statistics}, pages 492--518. Springer, 1992.

\bibitem{ironi2005colorization}
Revital Ironi, Daniel Cohen-Or, and Dani Lischinski.
\newblock Colorization by example.
\newblock In {\em Rendering techniques}, pages 201--210. Citeseer, 2005.

\bibitem{isola2017image}
Phillip Isola, Jun-Yan Zhu, Tinghui Zhou, and Alexei~A Efros.
\newblock Image-to-image translation with conditional adversarial networks.
\newblock In {\em Proceedings of the IEEE conference on computer vision and
  pattern recognition}, pages 1125--1134, 2017.

\bibitem{johnson2016perceptual}
Justin Johnson, Alexandre Alahi, and Li Fei-Fei.
\newblock Perceptual losses for real-time style transfer and super-resolution.
\newblock In {\em European conference on computer vision}, pages 694--711.
  Springer, 2016.

\bibitem{kingma2014adam}
Diederik~P Kingma and Jimmy Ba.
\newblock Adam: A method for stochastic optimization.
\newblock {\em arXiv preprint arXiv:1412.6980}, 2014.

\bibitem{kumar2021colorization}
Manoj Kumar, Dirk Weissenborn, and Nal Kalchbrenner.
\newblock Colorization transformer.
\newblock {\em arXiv preprint arXiv:2102.04432}, 2021.

\bibitem{larsson2016learning}
Gustav Larsson, Michael Maire, and Gregory Shakhnarovich.
\newblock Learning representations for automatic colorization.
\newblock In {\em European conference on computer vision}, pages 577--593.
  Springer, 2016.

\bibitem{levin2004colorization}
Anat Levin, Dani Lischinski, and Yair Weiss.
\newblock Colorization using optimization.
\newblock In {\em ACM SIGGRAPH 2004 Papers}, pages 689--694. 2004.

\bibitem{liao2017visual}
Jing Liao, Yuan Yao, Lu Yuan, Gang Hua, and Sing~Bing Kang.
\newblock Visual attribute transfer through deep image analogy.
\newblock {\em arXiv preprint arXiv:1705.01088}, 2017.

\bibitem{lin2017structured}
Zhouhan Lin, Minwei Feng, Cicero Nogueira~dos Santos, Mo Yu, Bing Xiang, Bowen
  Zhou, and Yoshua Bengio.
\newblock A structured self-attentive sentence embedding.
\newblock {\em arXiv preprint arXiv:1703.03130}, 2017.

\bibitem{liu2008intrinsic}
Xiaopei Liu, Liang Wan, Yingge Qu, Tien-Tsin Wong, Stephen Lin, Chi-Sing Leung,
  and Pheng-Ann Heng.
\newblock Intrinsic colorization.
\newblock In {\em ACM SIGGRAPH Asia 2008 papers}, pages 1--9. 2008.

\bibitem{lu2020gray2colornet}
Peng Lu, Jinbei Yu, Xujun Peng, Zhaoran Zhao, and Xiaojie Wang.
\newblock Gray2colornet: Transfer more colors from reference image.
\newblock In {\em Proceedings of the 28th ACM International Conference on
  Multimedia}, pages 3210--3218, 2020.

\bibitem{luan2007natural}
Qing Luan, Fang Wen, Daniel Cohen-Or, Lin Liang, Ying-Qing Xu, and Heung-Yeung
  Shum.
\newblock Natural image colorization.
\newblock In {\em Proceedings of the 18th Eurographics conference on Rendering
  Techniques}, pages 309--320, 2007.

\bibitem{mao2017least}
Xudong Mao, Qing Li, Haoran Xie, Raymond~YK Lau, Zhen Wang, and Stephen
  Paul~Smolley.
\newblock Least squares generative adversarial networks.
\newblock In {\em Proceedings of the IEEE international conference on computer
  vision}, pages 2794--2802, 2017.

\bibitem{morimoto2009automatic}
Yuji Morimoto, Yuichi Taguchi, and Takeshi Naemura.
\newblock Automatic colorization of grayscale images using multiple images on
  the web.
\newblock In {\em SIGGRAPH 2009: Talks}, pages 1--1. 2009.

\bibitem{parikh2016decomposable}
Ankur~P Parikh, Oscar T{\"a}ckstr{\"o}m, Dipanjan Das, and Jakob Uszkoreit.
\newblock A decomposable attention model for natural language inference.
\newblock {\em arXiv preprint arXiv:1606.01933}, 2016.

\bibitem{paszke2019pytorch}
Adam Paszke, Sam Gross, Francisco Massa, Adam Lerer, James Bradbury, Gregory
  Chanan, Trevor Killeen, Zeming Lin, Natalia Gimelshein, Luca Antiga, et~al.
\newblock Pytorch: An imperative style, high-performance deep learning library.
\newblock {\em arXiv preprint arXiv:1912.01703}, 2019.

\bibitem{pitie2007automated}
Fran{\c{c}}ois Piti{\'e}, Anil~C Kokaram, and Rozenn Dahyot.
\newblock Automated colour grading using colour distribution transfer.
\newblock {\em Computer Vision and Image Understanding}, 107(1-2):123--137,
  2007.

\bibitem{puzicha1997non}
Jan Puzicha, Thomas Hofmann, and Joachim~M Buhmann.
\newblock Non-parametric similarity measures for unsupervised texture
  segmentation and image retrieval.
\newblock In {\em Proceedings of IEEE Computer Society Conference on Computer
  Vision and Pattern Recognition}, pages 267--272. IEEE, 1997.

\bibitem{qu2006manga}
Yingge Qu, Tien-Tsin Wong, and Pheng-Ann Heng.
\newblock Manga colorization.
\newblock {\em ACM Transactions on Graphics (TOG)}, 25(3):1214--1220, 2006.

\bibitem{ramachandran2019stand}
Prajit Ramachandran, Niki Parmar, Ashish Vaswani, Irwan Bello, Anselm Levskaya,
  and Jonathon Shlens.
\newblock Stand-alone self-attention in vision models.
\newblock {\em arXiv preprint arXiv:1906.05909}, 2019.

\bibitem{reinhard2001color}
Erik Reinhard, Michael Adhikhmin, Bruce Gooch, and Peter Shirley.
\newblock Color transfer between images.
\newblock {\em IEEE Computer graphics and applications}, 21(5):34--41, 2001.

\bibitem{ren2015faster}
Shaoqing Ren, Kaiming He, Ross Girshick, and Jian Sun.
\newblock Faster r-cnn: Towards real-time object detection with region proposal
  networks.
\newblock {\em arXiv preprint arXiv:1506.01497}, 2015.

\bibitem{ronneberger2015u}
Olaf Ronneberger, Philipp Fischer, and Thomas Brox.
\newblock U-net: Convolutional networks for biomedical image segmentation.
\newblock In {\em International Conference on Medical image computing and
  computer-assisted intervention}, pages 234--241. Springer, 2015.

\bibitem{simonyan2014very}
Karen Simonyan and Andrew Zisserman.
\newblock Very deep convolutional networks for large-scale image recognition.
\newblock {\em arXiv preprint arXiv:1409.1556}, 2014.

\bibitem{tai2005local}
Yu-Wing Tai, Jiaya Jia, and Chi-Keung Tang.
\newblock Local color transfer via probabilistic segmentation by
  expectation-maximization.
\newblock In {\em 2005 IEEE Computer Society Conference on Computer Vision and
  Pattern Recognition (CVPR'05)}, volume~1, pages 747--754. IEEE, 2005.

\bibitem{vaswani2017attention}
Ashish Vaswani, Noam Shazeer, Niki Parmar, Jakob Uszkoreit, Llion Jones,
  Aidan~N Gomez, Lukasz Kaiser, and Illia Polosukhin.
\newblock Attention is all you need.
\newblock {\em arXiv preprint arXiv:1706.03762}, 2017.

\bibitem{vitoria2020chromagan}
Patricia Vitoria, Lara Raad, and Coloma Ballester.
\newblock Chromagan: adversarial picture colorization with semantic class
  distribution.
\newblock In {\em Proceedings of the IEEE/CVF Winter Conference on Applications
  of Computer Vision}, pages 2445--2454, 2020.

\bibitem{wang2020axial}
Huiyu Wang, Yukun Zhu, Bradley Green, Hartwig Adam, Alan Yuille, and
  Liang-Chieh Chen.
\newblock Axial-deeplab: Stand-alone axial-attention for panoptic segmentation.
\newblock In {\em European Conference on Computer Vision}, pages 108--126.
  Springer, 2020.

\bibitem{wang2018high}
Ting-Chun Wang, Ming-Yu Liu, Jun-Yan Zhu, Andrew Tao, Jan Kautz, and Bryan
  Catanzaro.
\newblock High-resolution image synthesis and semantic manipulation with
  conditional gans.
\newblock In {\em Proceedings of the IEEE conference on computer vision and
  pattern recognition}, pages 8798--8807, 2018.

\bibitem{wang2018non}
Xiaolong Wang, Ross Girshick, Abhinav Gupta, and Kaiming He.
\newblock Non-local neural networks.
\newblock In {\em Proceedings of the IEEE conference on computer vision and
  pattern recognition}, pages 7794--7803, 2018.

\bibitem{welsh2002transferring}
Tomihisa Welsh, Michael Ashikhmin, and Klaus Mueller.
\newblock Transferring color to greyscale images.
\newblock In {\em Proceedings of the 29th annual conference on Computer
  graphics and interactive techniques}, pages 277--280, 2002.

\bibitem{xiao2020example}
Chufeng Xiao, Chu Han, Zhuming Zhang, Jing Qin, Tien-Tsin Wong, Guoqiang Han,
  and Shengfeng He.
\newblock Example-based colourization via dense encoding pyramids.
\newblock In {\em Computer Graphics Forum}, volume~39, pages 20--33. Wiley
  Online Library, 2020.

\bibitem{xu2015empirical}
Bing Xu, Naiyan Wang, Tianqi Chen, and Mu Li.
\newblock Empirical evaluation of rectified activations in convolutional
  network.
\newblock {\em arXiv preprint arXiv:1505.00853}, 2015.

\bibitem{xu2009efficient}
Kun Xu, Yong Li, Tao Ju, Shi-Min Hu, and Tian-Qiang Liu.
\newblock Efficient affinity-based edit propagation using kd tree.
\newblock {\em ACM Transactions on Graphics (TOG)}, 28(5):1--6, 2009.

\bibitem{xu2020stylization}
Zhongyou Xu, Tingting Wang, Faming Fang, Yun Sheng, and Guixu Zhang.
\newblock Stylization-based architecture for fast deep exemplar colorization.
\newblock In {\em Proceedings of the IEEE/CVF Conference on Computer Vision and
  Pattern Recognition}, pages 9363--9372, 2020.

\bibitem{yatziv2006fast}
Liron Yatziv and Guillermo Sapiro.
\newblock Fast image and video colorization using chrominance blending.
\newblock {\em IEEE transactions on image processing}, 15(5):1120--1129, 2006.

\bibitem{zhang2019deep}
Bo Zhang, Mingming He, Jing Liao, Pedro~V Sander, Lu Yuan, Amine Bermak, and
  Dong Chen.
\newblock Deep exemplar-based video colorization.
\newblock In {\em Proceedings of the IEEE/CVF Conference on Computer Vision and
  Pattern Recognition}, pages 8052--8061, 2019.

\bibitem{zhang2019self}
Han Zhang, Ian Goodfellow, Dimitris Metaxas, and Augustus Odena.
\newblock Self-attention generative adversarial networks.
\newblock In {\em International conference on machine learning}, pages
  7354--7363. PMLR, 2019.

\bibitem{zhang2016colorful}
Richard Zhang, Phillip Isola, and Alexei~A Efros.
\newblock Colorful image colorization.
\newblock In {\em European conference on computer vision}, pages 649--666.
  Springer, 2016.

\bibitem{zhang2017real}
Richard Zhang, Jun-Yan Zhu, Phillip Isola, Xinyang Geng, Angela~S Lin, Tianhe
  Yu, and Alexei~A Efros.
\newblock Real-time user-guided image colorization with learned deep priors.
\newblock {\em arXiv preprint arXiv:1705.02999}, 2017.

\end{thebibliography}
}

\end{document}